\pdfoutput=1

\documentclass[]{spie}  

\usepackage{amsmath,amsfonts,amssymb}
\usepackage{graphicx}
\usepackage[colorlinks=true, allcolors=blue]{hyperref}
\usepackage{gensymb}
\usepackage{siunitx}
\usepackage{blindtext}
\usepackage{longtable} 
\usepackage{booktabs}

\title{Characterization of low-loss hydrogenated amorphous silicon films for superconducting resonators}

\author[a]{B.T. Buijtendorp}
\author[a,b]{J. Bueno}
\author[a]{D.J. Thoen}
\author[b]{V. Murugesan}
\author[a]{P.M. Sberna}
\author[a,b]{J.J.A.~Baselmans}
\author[a]{S. Vollebregt}
\author[a,c]{A. Endo}

\affil[a]{Faculty of Electrical Engineering, Mathematics and Computer Science, Delft University of Technology, Delft, The Netherlands}
\affil[b]{SRON -- Netherlands Institute for Space Research, Utrecht, The Netherlands}
\affil[c]{Kavli Institute of NanoScience, Faculty of Applied Sciences, Delft University of Technology, Delft, The Netherlands}

\authorinfo{B.T.B.: E-mail: b.t.buijtendorp@tudelft.nl}

\begin{document} 
\maketitle

\begin{abstract}
Superconducting resonators used in millimeter-submillimeter astronomy would greatly benefit from deposited dielectrics with a small dielectric loss. We deposited hydrogenated amorphous silicon films using plasma-enhanced chemical vapor deposition, at substrate temperatures of 100\degree C, 250\degree C and 350\degree C.  The measured void volume fraction, hydrogen content, microstructure parameter, and bond-angle disorder are negatively correlated with the substrate temperature. All three films have a loss tangent below $10^{-5}$ for a resonator energy of $10^{5}$ photons, at 120 mK and 4--7 GHz.  This makes these films promising for microwave kinetic inductance detectors and on-chip millimeter-submilimeter filters.
\end{abstract}

\keywords{Kinetic Inductance Detectors, Millimeter-wave, Submillimeter-wave, Amorphous Silicon, Spectrometer, Filter Bank, Two-level Systems, Dielectric Loss}

\section{INTRODUCTION}
The integrated superconducting spectrometer (ISS)\cite{wheeler_superspec_2018, cataldo_second-generation_2018, endo_first_2019} is a novel astronomical instrument that promises ultra-wideband, high-sensitivity, and three-dimensional imaging spectrometry in the millimeter-submillimeter (mm-submm) band. ISSs enable large arrays of spectroscopic pixels, because they are intrinsically scalable due to the compact on-chip filterbank for signal dispersion, and because they allow for frequency-domain multiplexed readout of the microwave kinetic inductance detectors (MKIDs) \cite{day_broadband_2003}. 

Currently, the DESHIMA\cite{endo_first_2019} ISS has a coplanar waveguide (CPW) signal line coupled to planar filters. This planar technology has the advantage that it can easily be fabricated on top of crystalline dielectrics, which exhibit lower dielectric losses than their amorphous counterparts \cite{oconnell_microwave_2008}. Unfortunately this planar filterbank suffers from large radiation losses \cite{Endo2019filterbank, Hahnle2020}, which could be eliminated by using microstrip filters. However, the microstrip filters cannot easily be fabricated without using a deposited dielectric \cite{Endo2013, wheeler_superspec_2018}. 

The transmission $|S_{31}|^2$ through a single isolated mm-submm filter to its detector is determined by the filter's resolution $R = F/\Delta F$ and its internal quality factor $Q_i$ \cite{Endo2013, Hailey-Dunsheath2014}:

\begin{equation}
	|S_{31}|^2 = \frac{1}{2}\left(1 - \frac{R}{Q_i} \right)^2
.\end{equation}

\noindent The ISSs DESHIMA and SuperSpec\cite{wheeler_superspec_2018} both aim for $R\sim$ 300--500. The $Q_i$ is dominated by dielectric losses due to two-level systems (TLSs), and currently it remains a challenge to deposit a dielectric with a sufficiently large $Q_i$. To date, the largest reported $Q_i$ for mm-submm microstrip filters is 1440 and has been achieved with an amorphous silicon nitride ($\mathrm{SiN}_x$) dielectric \cite{Hailey-Dunsheath2014}. For $R=500$, this is  equivalent to $|S_{31}|^2 = 21\%$. To put this into perspective: A $Q_i = 5000$ would result in $|S_{31}|^2 = 40\%$. Clearly, a significant improvement can be achieved by depositing a more transparent dielectric.

Apart from its application in experimental astronomy, an improved dielectric will also be beneficial to other fields, for example in superconducting quantum computation \cite{oconnell_microwave_2008}. Additionally, since the microsopic origin of the TLSs remains unknown \cite{muller_towards_2019}, a better understanding of the TLSs is interesting from a fundamental perspective. 

TLSs are sucessfully modelled by the standard tunneling model (STM) \cite{phillips_tunneling_1972}. In the STM the TLS density of states is frequency independent, and this make it reasonable to assume that a decreased microwave dielectric loss can be translated to increased performance for the mm-submm filters. However, this frequency independence should be confirmed experimentally. Currently, the best low-power microwave loss tangents have been measured for hydrogenated amorphous silicon (a-Si:H) films \cite{mazin_thin_2010, oconnell_microwave_2008}. Using lumped element resonators, a low-power microwave $\tan{\delta} \sim 10^{-5}$ was observed, 5--10 times smaller than for $\si{SiN}_x$ films \cite{oconnell_microwave_2008} measured at similar microwave frequencies and readout powers.

Due to an observed correlation between TLS density and atomic silicon density, it has been proposed that the TLSs in a-Si:H are related to voids in the dielectric \cite{queen_two-level_2015}. It was found that a very low void volume fraction can be achieved by depositing at elevated substrate temperatures $T_{\si{sub}}$\cite{queen_two-level_2015}. Recently, it has been observed that although the TLS-induced internal friction correlates with the atomic silicon density, the TLS-induced loss tangent instead correlates with the dangling bond density \cite{Molina-Ruiz2020}.

In this work we investigate the effect of $T_{\si{sub}}$ on the microstructural and compositional properties of a-Si:H, and we study how these properties relate to the dielectric loss at 4--7 GHz and at 120 mK. For this purpose we deposited a-Si:H films using plasma-enhanced chemical vapor deposition (PECVD) at three different substrate temperatures, and we measured their void volume fraction, hydrogen content, microstructure parameter, bond-angle disorder and infrared (IR) refractive index.

\section{Film deposition}
We deposited the a-Si:H films using PECVD with an Oxford Plasmalab 80 Plus. The substrate temperatures $T_{\si{sub}}$ during the PECVD processes were 100\degree C, 250\degree C and 350\degree C. The $T_{\si{sub}}$ was the only parameter that was varied. The deposition parameters are listed in Appendix \ref{app:depo}. We used crystalline silicon (c-Si) substrates with (100) crystal orientation. Further substrate details and the wafer preparation details depend on the subsequent experiment, and are described in the methods of each experiment. 
\section{Room temperature characterization}
\subsection{Hydrogen content, microstructure parameter, infrared refractive index}
\begin{figure} [ht]
\begin{center}
\begin{tabular}{c} 
\includegraphics[height=9cm]{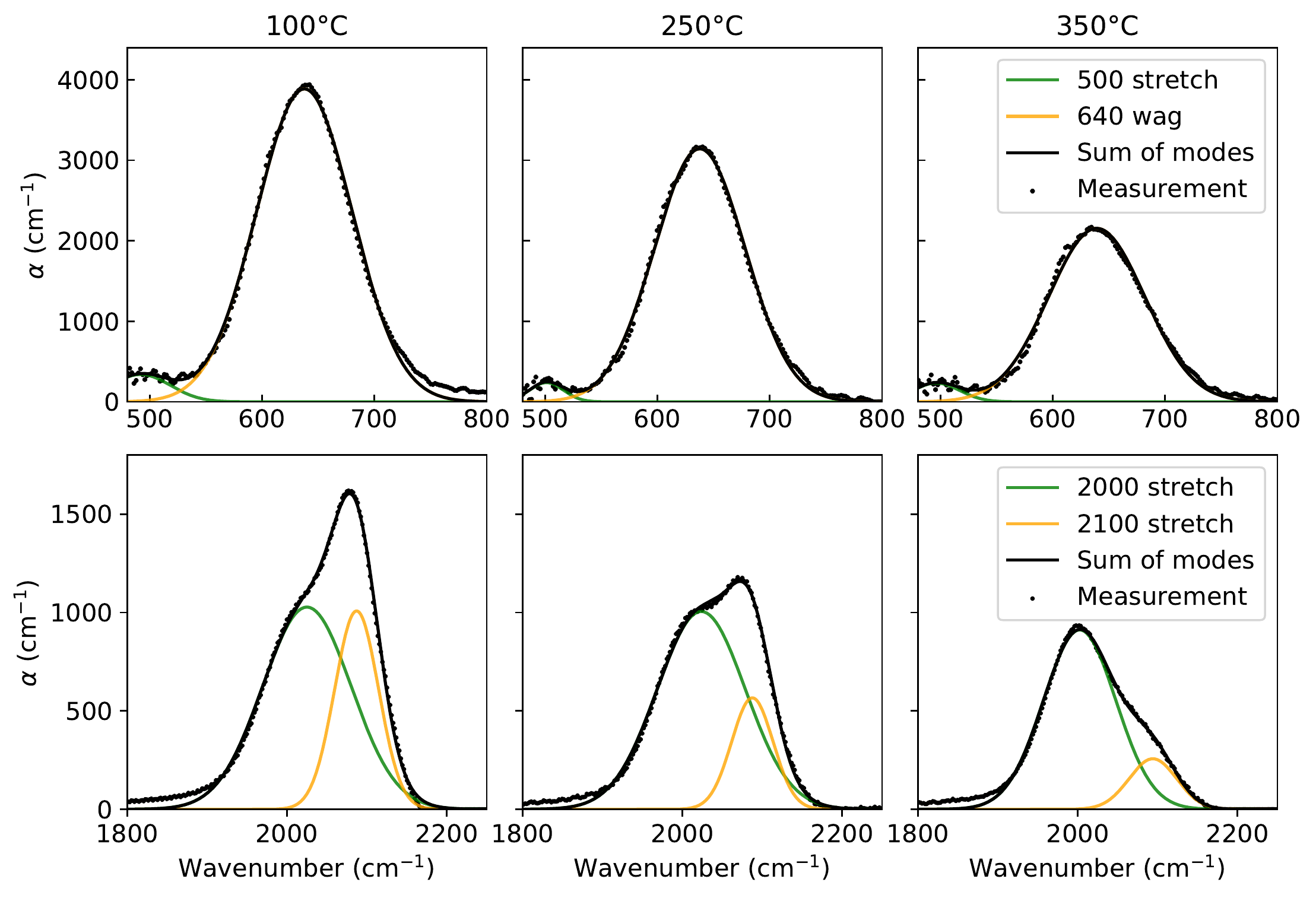}
\end{tabular}
\end{center}
\caption[Test]{\label{fig:abs}The absorption coefficients $\alpha$ that we measured using FTIR, as a function of wavenumber $\tilde{\nu}$. Each column shows the results for a particular film, that was deposited at the substrate temperature $T_{\si{sub}}$ that is displayed on top. The top row shows the absorption near the wagging mode at 640 $\si{cm^{-1}}$ (referred to as 640 wag), from which we calculated the hydrogen content $C_{\si{H}}$. The bottom row shows the absorption near the stretching modes at 2000 and 2100 $\si{cm^{-1}}$ (referred to as 2000 stretch and 2100 stretch), from which we calculated the microstructure parameter $R^*$.} 
\end{figure} 

\begin{figure} [ht]
\begin{center}
\begin{tabular}{c} 
\includegraphics[height=6cm]{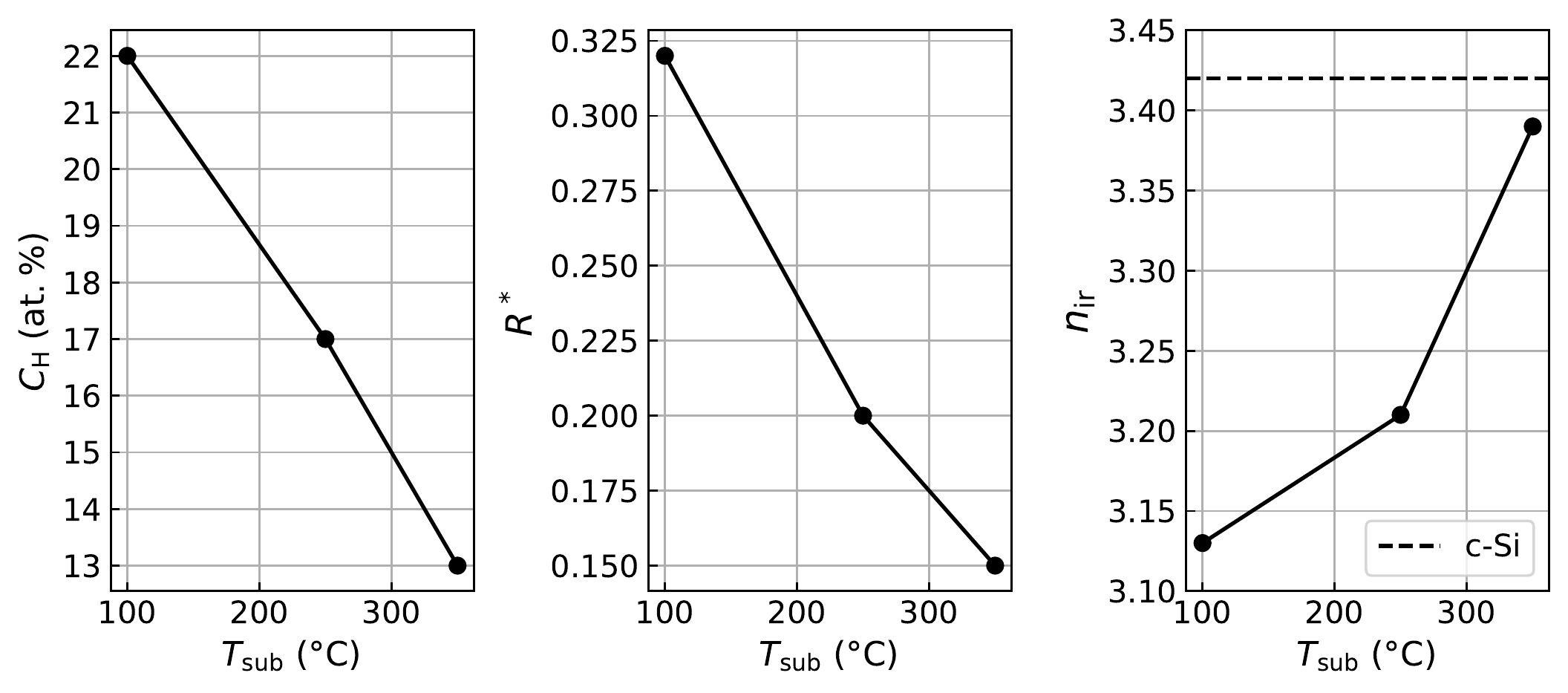}
\end{tabular}
\end{center}
\caption[Test]{\label{fig:ftirres}The hydrogen contents $C_{\si{H}}$, microstructure parameters $R^*$, and infrared refractive indices $n_{\si{ir}}$ of the three films, that we derived from the FTIR data. The films were deposited at the substrate temperature $T_{\si{sub}}$. The dashed line in the right-most figure represents the literature value of the $n_{\si{ir}}$ of c-Si \cite{langford_infrared_1992}.} 
\end{figure}
The microstructure of a-Si:H is largely governed by the occurrence of $\si{SiH}$ and $\si{SiH_2}$ configurations in the material \cite{ouwens_hydrogen_1996, smets_vacancies_2003, beyer_voids_2012}. The $\si{SiH}$ configurations reside mostly in small vacancies, corresponding to up to three missing silicon (Si) atoms\cite{smets_vacancies_2003}. This is in contrast to the $\si{SiH_2}$ configurations that exist mostly on the surface of voids with a radius of 1--4 nm, corresponding to $\si{10^2}$ to $\si{10^4}$ missing Si atoms\cite{smets_vacancies_2003}. We used Fourier-transform infrared (FTIR) spectrocopy in transmission mode to measure the microstructure parameter $R^*$, that quantifies the relative amount of $\si{SiH}$ and $\si{SiH_2}$ bonds, as defined in Eq. \eqref{eq:micro}. Using this method we also obtained the total hydrogen content $C_{\si{H}}$ of the films, and their infrared refractive indices $n_{\si{ir}}$. 

For the FTIR measurements we deposited a-Si:H films on double side polished (DSP) p-type c-Si substrates with $\rho > 1 \ \si{k}\ohm$ \si{cm}. The substrates were dipped in 1\% hydrofluoric acid for one minute prior to film deposition.

By fitting the FTIR transmission data to a transfer-matrix method (TMM) calculation\cite{Byrnes2018} we obtained the $n_{\si{ir}}$ and the absorption coefficients $\alpha$ of the a-Si:H films. In the fitting we used a non-dispersive $n_{\si{ir}}$. We also measured the $n_{\si{ir}}$ and $\alpha$ of the bare c-Si substrate, and used this to model the substrate in the calculation.

The measured $\alpha$ and their Gaussian fits are plotted in Fig. \ref{fig:abs}. With increasing $T_{\si{sub}}$, we observe a decrease in intensity of the wagging mode near 640~\si{cm^{-1}} (referred to as 640 wag), indicating a decrease of the total hydrogen content\cite{ouwens_hydrogen_1996, langford_infrared_1992}. We also observe an absorption peak near 500 \si{cm^{-1}} that can be attributed to a stretching mode\cite{maley_critical_1992} (referred to as 500 stretch). Also with increasing $T_{\si{sub}}$, we observe a decrease in intensity of the stretching mode near 2100 \si{cm^{-1}}, indicating a decrease in the amount of $\mathrm{SiH_2}$ bonds.

To calculate the films' hydrogen contents $C_{\si{H}}$ and their microstructure parameters $R^*$ from the $\alpha$ spectra we numerically calculated the integrated absorptions $I_x$: 
\begin{equation}
	I_x = \int \frac{\alpha_x(\tilde{\nu})}{\tilde{\nu}}d\tilde{\nu}
,\end{equation}
where $x$ denotes the center wavenumber of the Gaussian absorption peak that is being integrated. From $I_{\si{640}}$ we calculated $C_{\si{H}}$ in atomic percent (at. \%) \cite{ouwens_hydrogen_1996, street_ra_hydrogenated_1991, langford_infrared_1992, smets_vacancies_2003}:
\begin{equation}
	C_H = \frac{A_{\si{640}}I_{\si{640}}}{A_{\si{640}}I_{\si{640}} + N_{\si{Si}}}
,\end{equation}
where $A_{\si{640}}$ is a proportionality constant that is an inverse cross-section for photon absorption at this wavenumber. We used the value $A_{\si{640}}$ = $2.1 \cdot 10^{19}$ $\si{cm^{-2}}$ \cite{langford_infrared_1992}. For $N_{\si{Si}}$, the atomic density of silicon, we used the value $\si{5 \cdot 10^{22}}~\si{cm^{-3}}$.
The modes near 2000 \si{cm^{-1}} and 2100 \si{cm^{-1}} can be attributed to $\si{SiH}$ and $\si{SiH_2}$, respectively \cite{ouwens_hydrogen_1996}. The fraction of Si atoms that are bonded as $\si{SiH_3}$ is negligible below hydrogen concentrations of 40\% \cite{mui_optical_1988}. $R^*$ was defined as \cite{ouwens_hydrogen_1996}:
\begin{equation}\label{eq:micro}
	R^* = \frac{I_{\si{2100}}}{I_{\si{2100}} + I_{\si{2000}}}
.\end{equation}

The results for the hydrogen content $C_{\si{H}}$, microstructure parameter $R^*$, and infrared refractive index $n_{\si{ir}}$ are plotted in Fig. \ref{fig:ftirres}. We observe that $C_{\si{H}}$ decreases monotonically with increasing $T_{\si{sub}}$. Furthermore,  the microstructure parameter $R^*$ decreases monotonically with increasing $T_{\si{sub}}$, indicating that a smaller fraction of the hydrogen is bonded in $\si{SiH_2}$ configurations that reside mostly on the surface of voids\cite{smets_vacancies_2003}. Additionally, we observe that $n_{\si{ir}}$ increases monotonically with increasing $T_{\si{sub}}$, indicative of an increase in film density.
\subsection{Bond-angle disorder}
\begin{figure} [ht]
\begin{center}
\begin{tabular}{c} 
\includegraphics[height=6cm]{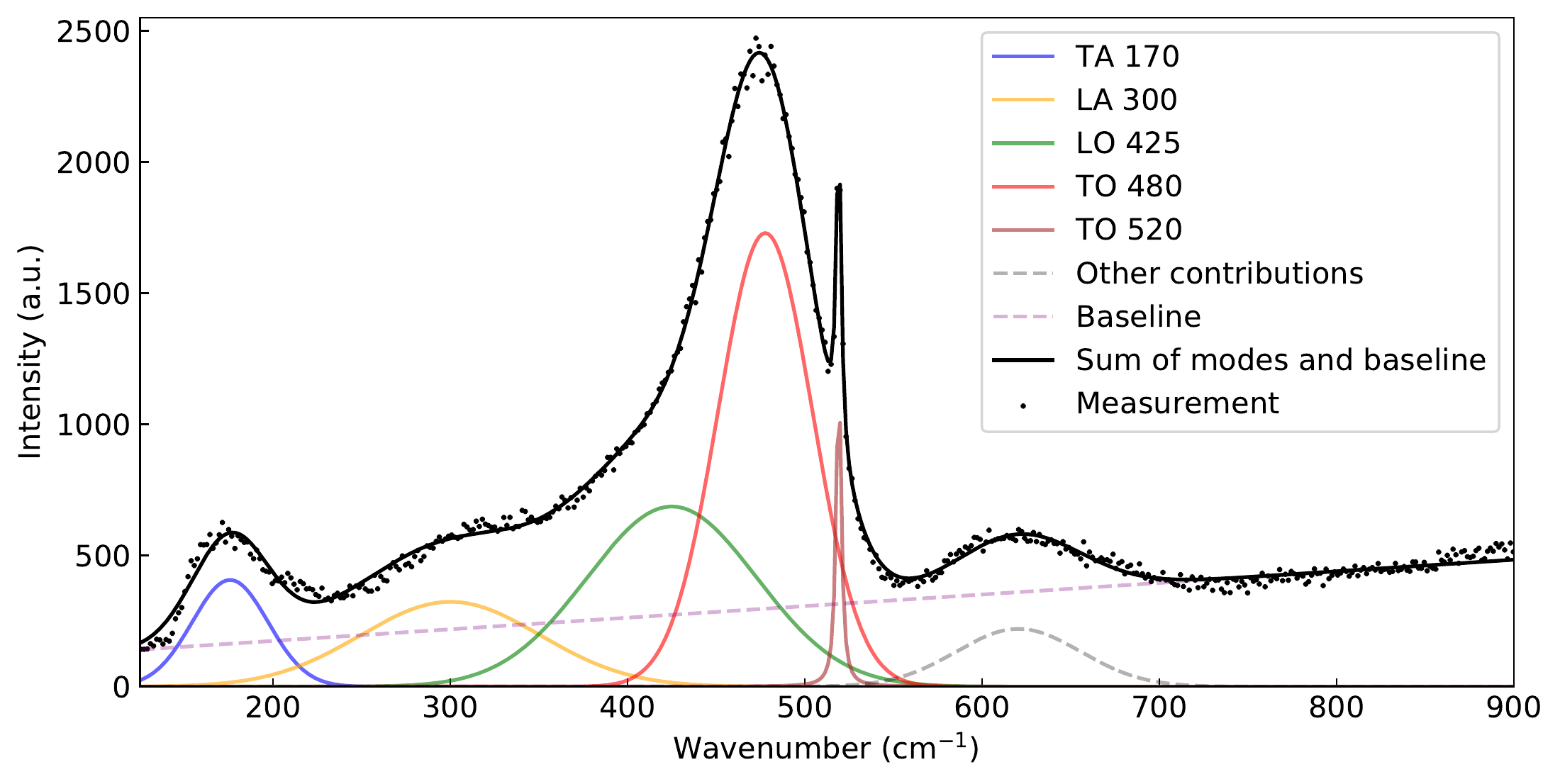}
\end{tabular}
\end{center}
\caption[Test]{\label{fig:raman}The Raman spectroscopy measurement of the a-Si:H film that was deposited at a $T_{\si{sub}}$ of 100\degree C. The plot serves to show the modes that are fitted to the measurement data. We calculated the bond-angle disorder $\Delta\theta$ from the position of the TO 480 mode.} 
\end{figure} 
\begin{figure} [ht]
\begin{center}
\begin{tabular}{c} 
\includegraphics[height=5cm]{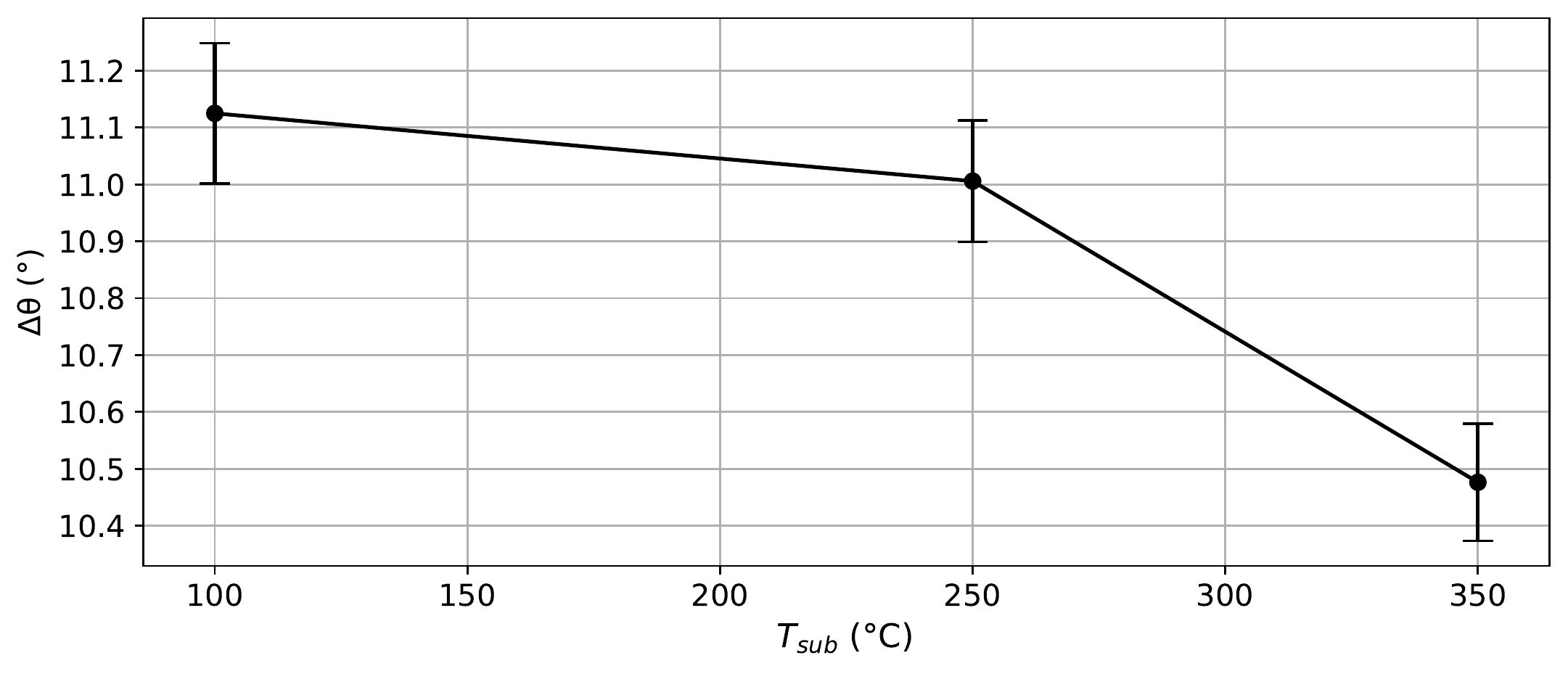}
\end{tabular}
\end{center}
\caption[Test]{\label{fig:ramanres}The bond-angle disorder $\Delta\theta$ of the three a-Si:H films that we derived from the Raman spectroscopy measurements. The films were deposited at the substrate temperature $T_{\si{sub}}$. The error bars represent the standard deviations of $\Delta\theta$ that we estimated from the least-squares fit.} 
\end{figure} 
\noindent Films of a-Si:H exhibit bond-angle disorder $\Delta\theta$ in their silicon network, defined as the root mean square deviation from the tetrahedral bond angle of 109.5$\degree$. We measured  $\Delta\theta$ using Raman spectroscopy with a 514 nm laser \cite{vink_raman_2001}:
\begin{equation}\label{eq:raman}
	\Delta\theta = \frac{505.5-\tilde{\nu}_{\si{TO}}}{2.5}
,\end{equation}
where $\tilde{\nu}_{\si{TO}}$ is the center wavenumber of the transverse-optic (TO) mode near 480 \si{cm^{-1}} (TO 480). 

For Raman spectroscopy we deposited the a-Si:H films on single side polished (SSP) n-type substrates with a resistivity $\rho$ of 1--5 $\ohm$ \si{cm}, and with a 101-nm thick thermal oxide layer.

In Fig. \ref{fig:raman} we show the Raman measurement of the 100 \degree C film, and the fitted Gaussians. Apart from the TO 480 mode, we observe a transverse-acoustic (TA) mode near 170 \si{cm^{-1}} (TA 170) \cite{lockwood_optical_1993}, a longitudinal-acoustic mode near 300 \si{cm^{-1}} (LA 300) \cite{brodsky_infrared_1977}, a longitudinal-optic mode near 425 \si{cm^{-1}} (LO 425), and at 520 \si{cm^{-1}} the TO mode of the c-Si substrate, fitted with a Lorentzian \cite{vink_raman_2001, smit_determining_2003}. The peak near 620 \si{cm^{-1}} can be attributed to a variety of modes \cite{brodsky_infrared_1977, smit_determining_2003, 	mishra_first-_2001}. In our fit we fixed the peak position of the LA 300 mode at 300 \si{cm^{-1}}.

The measured $\Delta\theta$ of the three films are plotted in Fig. \ref{fig:ramanres}. We observe a monotonic decrease of $\Delta\theta$ with increasing $T_{\si{sub}}$.
\subsection{Void volume fraction}
We determined the films' void volume fractions $f_{\si{v}}$ using variable-angle spectroscopic ellipsometry. The measured change in polarization was fitted to an optical model that includes a native oxide layer, the a-Si:H film, a thermal oxide layer, and the c-Si substrate. For this we used the commercial software CompleteEASE\cite{CompleteEASE}. 

For ellipsometry we deposited the a-Si:H films on single side polished (SSP) n-type substrates with a resistivity $\rho$ of 1--5 $\ohm$ \si{cm}, with a 101-nm thick thermal oxide layer.

\begin{figure} [ht]
\begin{center}
\begin{tabular}{c} 
\includegraphics[height=5cm]{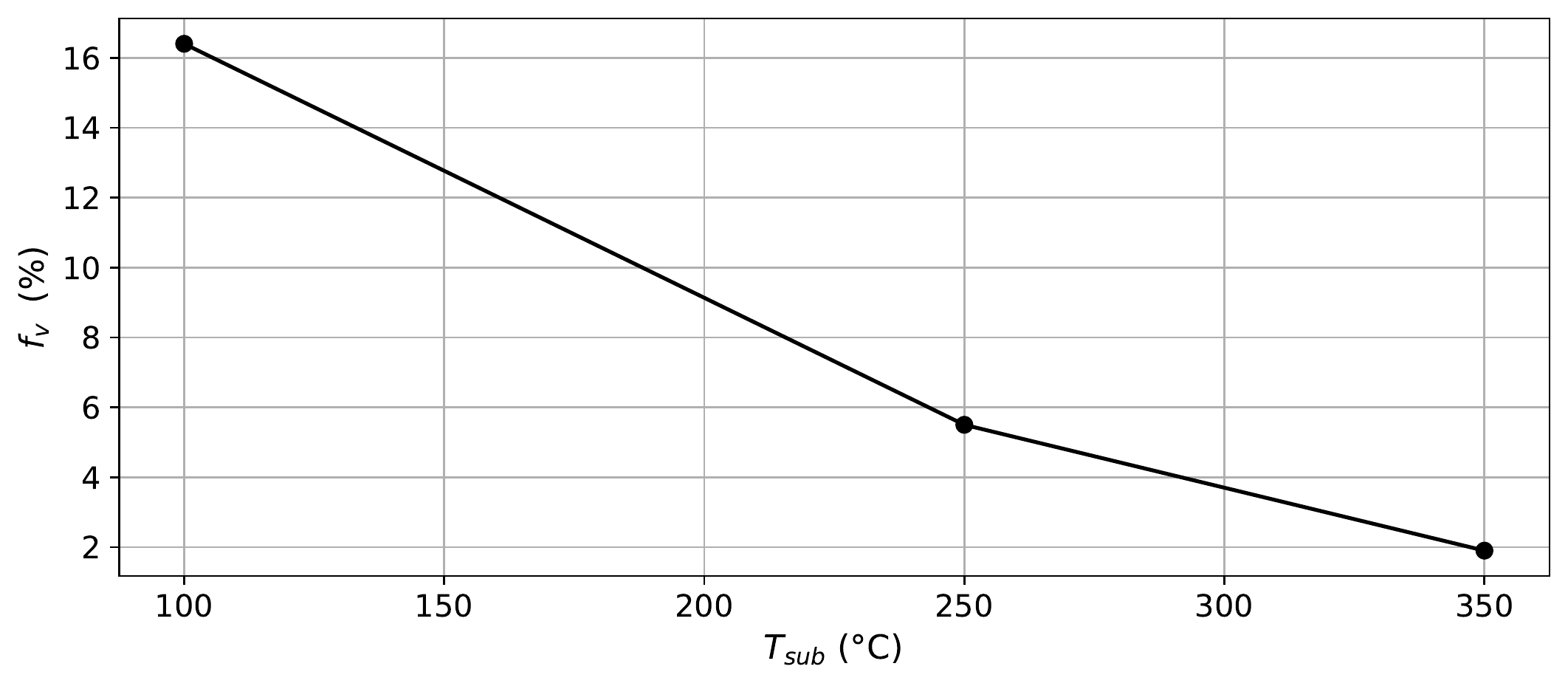}
\end{tabular}
\end{center}
\caption[Test]{\label{fig:ellipsres}The void volume fractions $f_{\si{v}}$ of the three a-Si:H films, that we derived from the ellipsometry data by modeling the a-Si:H layer as a composite material of a-Si and spherical voids using the Bruggeman effective medium approximation. The films were deposited at the substrate temperature $T_{\si{sub}}$.} 
\end{figure}

We obtained $f_{\si{v}}$ from the fit by using the Bruggeman effective medium approximation to model the a-Si:H film as a composite material consisting of a-Si and spherical voids. The details of the host material in this model are given in Appendix \ref{app:EMA}. The results for $f_{\si{v}}$ are plotted in Fig. \ref{fig:ellipsres}. We observe that $f_{\si{v}}$ decreases monotonically with increasing $T_{\si{sub}}$.

\section{Loss tangent measurements}   
\subsection{Setup}
\begin{figure} [ht]
\begin{center}
\begin{tabular}{c} 
\includegraphics[height=3.5cm]{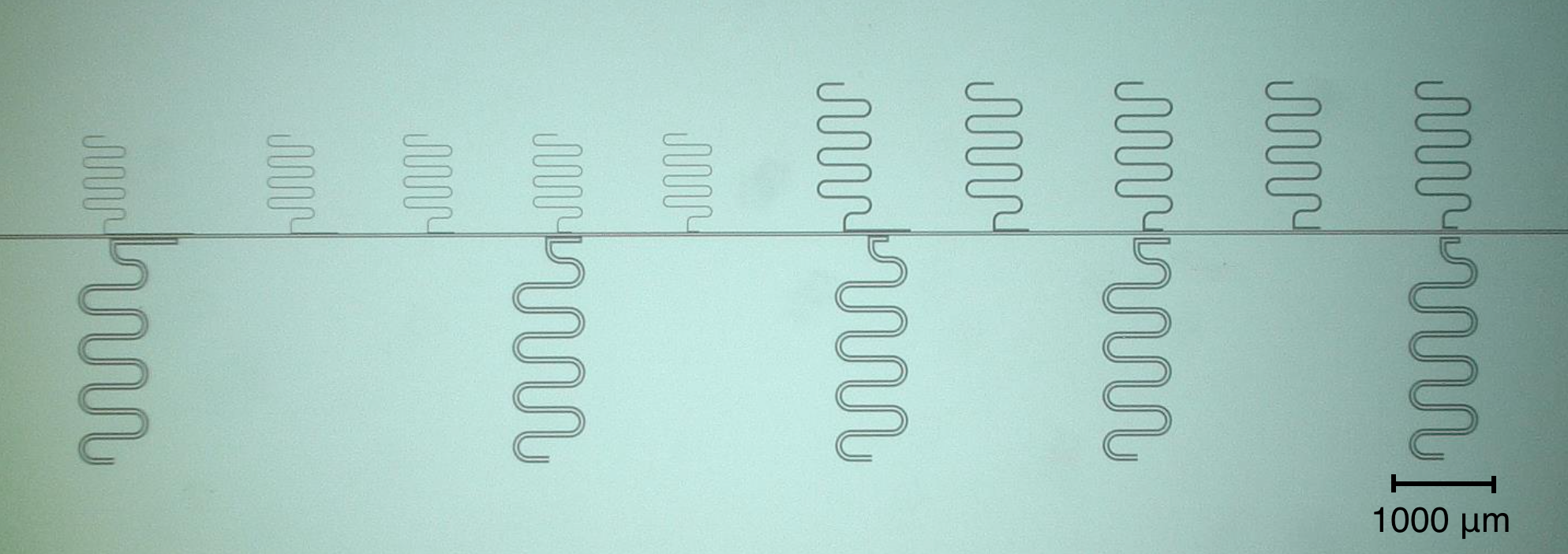}
\end{tabular}
\end{center}
\caption[Test]{\label{fig:chip}Micrograph of the superconducting chip. There are three groups of five quarter-wavelength CPW resonators, each group has a different CPW geometry (listed in Table \ref{tab:ff}). Within each group, the resonators vary in the coupling quality factor $Q_c$. A close-up micrograph of the resonator in the bottom-right of this figure is displayed in Fig. \ref{fig:chip2}.} 
\end{figure} 
\begin{figure} [ht]
\begin{center}
\begin{tabular}{c} 
\includegraphics[height=5cm]{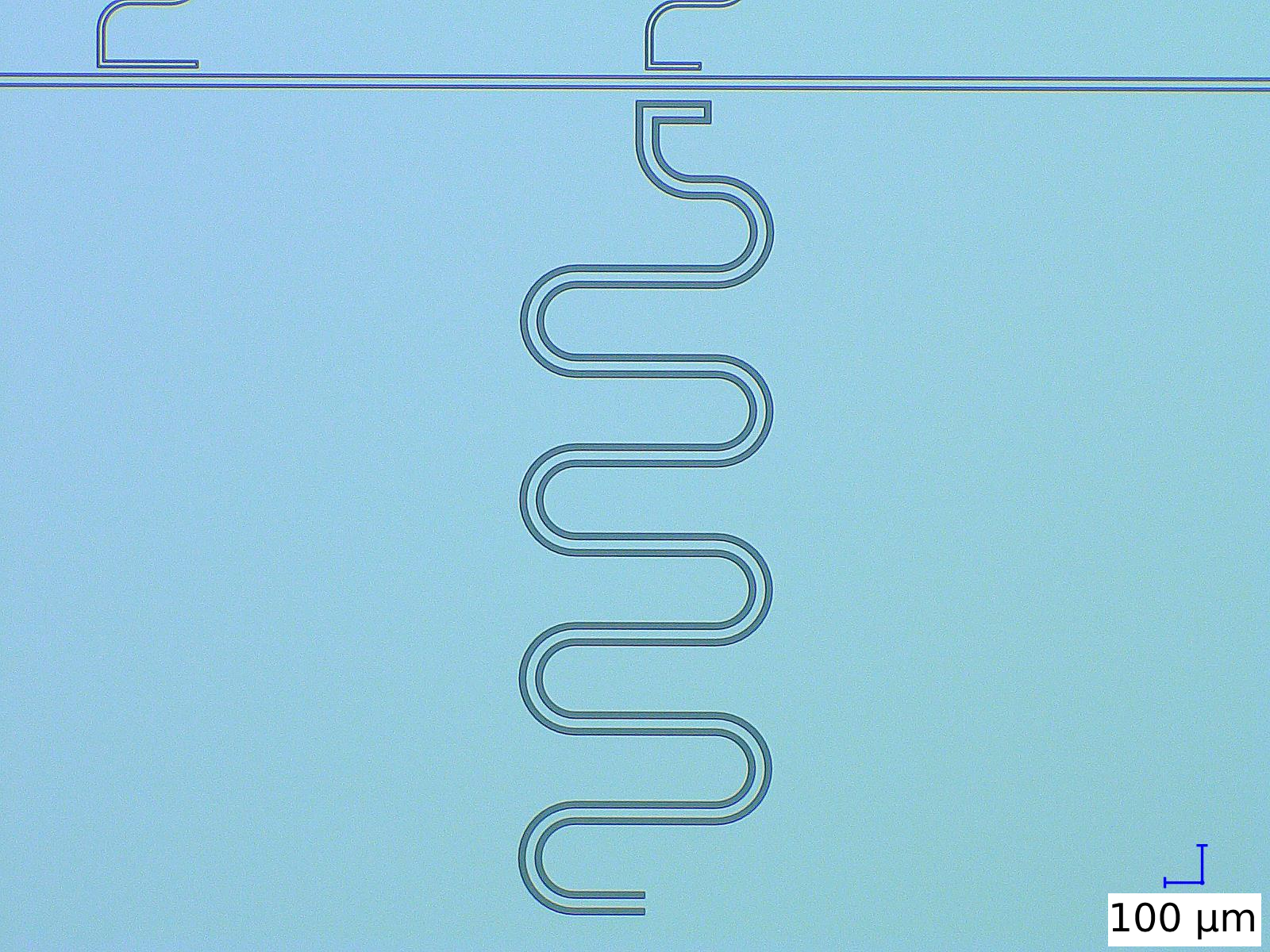}
\end{tabular}
\end{center}
\caption[Test]{\label{fig:chip2}Micrograph of one of the quarter-wavelength CPW resonators (bottom-right in Fig.\ref{fig:chip}). This resonator has a slot width $s$ of 18 $\si{\mu m}$ and a center line width $c$ of 27 $\si{\mu m}$.}
\end{figure} 
\begin{figure} [ht]
\begin{center}
\begin{tabular}{c} 
\includegraphics[height=5cm]{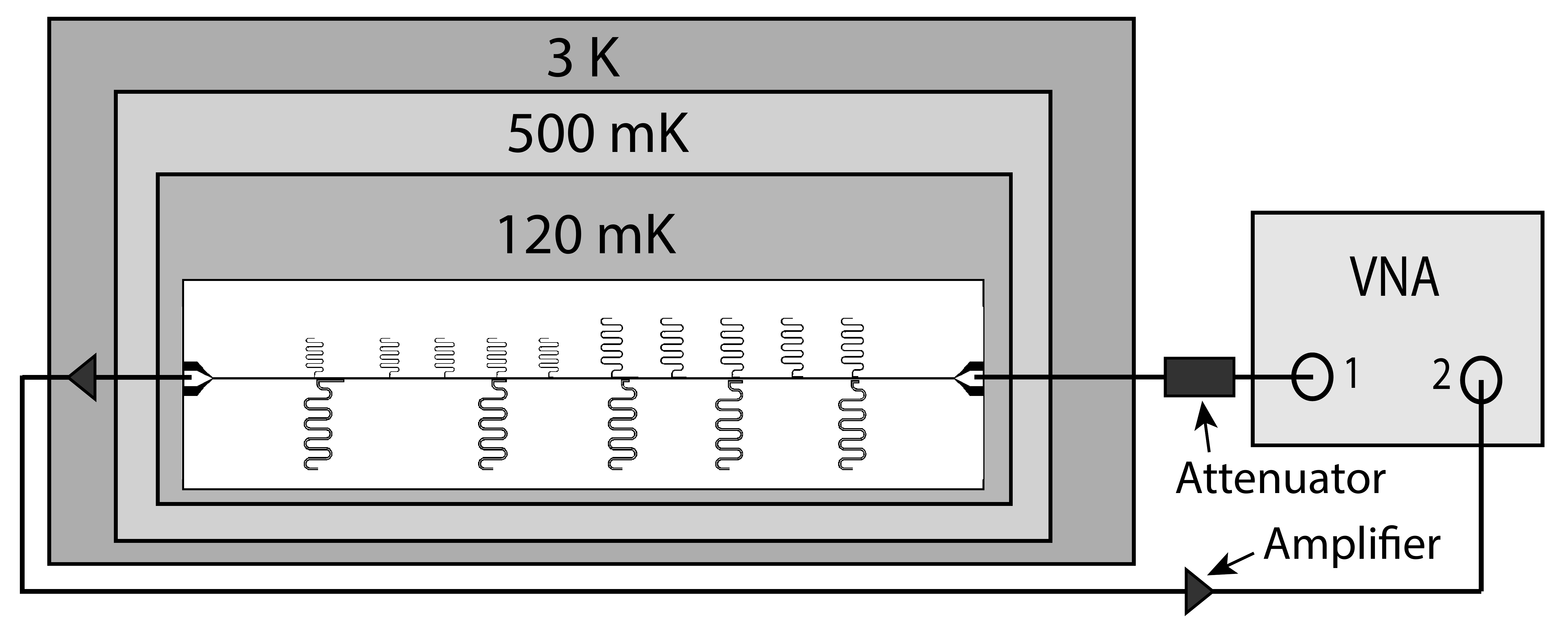}
\end{tabular}
\end{center}
\caption[Test]{\label{fig:cryosetup}Illustration of the cryogenic measurement setup. The two ports of the chip are connected to a vector network analyzer (VNA). The chip is cooled down to 120 mK with an adiabatic demagnetization refrigerator (ADR). The microwave signal from the VNA is attenuated before entering the chip. The signal is amplified at the 3 K stage and at room temperature. A detailed discussion of this measurement setup can be found in De Visser's PhD thesis \cite{DeVisser2014}.}
\end{figure} 
\noindent We measured the cryogenic microwave loss tangent $\tan{\delta}$ of the a-Si:H films at 4--7 GHz and at 120 mK. We used a superconducting chip with 109 nm thick aluminium quarter-wavelength CPW resonators. These were fabricated on top of the a-Si:H films using photolithography. For reference we also fabricated a chip directly on top of the c-Si substrate. 

For the loss tangent measurements we deposited a-Si:H films on DSP intrinsic c-Si substrates with $\rho > 10 \ \si{k}\ohm$ \si{cm}. The substrates were dipped in 1\% hydrofluoric acid for one minute prior to film deposition.

The chip is displayed in Fig. \ref{fig:chip} and features a CPW readout line with a slot width $s$ of 8 $\si{\mu m}$ and a center line width $c$ of 20 $\si{\mu m}$. the readout line is capacitively coupled to 15 resonators. As can be seen in Table \ref{tab:ff}, the resonators are divided into three groups, that each have a different CPW geometry ($s$ and $c$). The five resonators within each group were designed to have different coupling quality factors $Q_c$. For our results we use the data of one resonator per geometry group, because these have the smallest fitting errors. These are the resonators with the largest $Q_c \sim 3\cdot 10^5$.  

In Table \ref{tab:ff} we display the filling fractions $p$ of the dielectric films, i.e. the fraction of the resonator's electrostatic energy that is stored inside the dielectric \cite{gao_physics_2008}. The $p$ were calculated in Sonnet \cite{sonnet_user_guide}, where we used the film thicknesses $t_{\si{f}}$ that were measured using ellipsometry, and the microwave dielectric constants ($\epsilon_{\si{mw}}$) were estimated from the IR refractive indices $n_{\si{ir}}$ using Eq. \eqref{eq:est_mw}.

\begin{equation}
\label{eq:est_mw}
	\epsilon_{\si{mw}} \approx \left( \frac{n_{\si{ir}}}{n_{\si{cSi, ir}}}\right)^2 \epsilon_{\si{cSi, mw}} 	
.\end{equation}

To measure $\tan{\delta}$, the scattering parameter as a function of frequency $S_{\si{21}}(f)$ was measured using a vector network analyzer (VNA).  Figure \ref{fig:cryosetup} provides a schematic overview of the setup. 

\begin{table}[!htb]
\centering
\begin{tabular}{lccc}
             & \multicolumn{3}{c}{CPW s-c-s ($\si{\mu m}$)}\\ \cmidrule{2-4}
Material         & 2-3-2 & 6-9-6 & 18-27-18\\ \midrule
c-Si ($\epsilon_{\si{mw}}$ = 11.44)     & 0.920 & 0.920 & 0.920\\
a-Si 100 ($\epsilon_{\si{mw}} \approx$ 9.5)  & 0.208 & 0.101 & 0.044\\
a-Si 250 ($\epsilon_{\si{mw}} \approx$ 10.1)  & 0.197 & 0.095 & 0.042\\
a-Si 350 ($\epsilon_{\si{mw}} \approx$ 11.2)  & 0.214 & 0.103 & 0.045\\
\\
\end{tabular}
\caption{The filling fractions $p$ of the a-Si:H films that we calculated using Sonnet. In the case of c-Si, the $p$ is calculated for the substrate. The $\epsilon_{\si{mw}}$ of the a-Si:H films were estimated from their infrared values, as shown in Eq. \eqref{eq:est_mw}.} 
\label{tab:ff}
\end{table}

\subsection{Calculation of the loss tangent}
\label{sec:deter}
We determined the loss tangent $\tan{\delta}$ by fitting the loaded quality factor $Q$, the minimum of the resonance dip $S_{\si{21,min}}$, and the resonance frequency $f_{\si{r}}$ to the squared magnitude of the measured $S_{\si{21}}(f)$:
\begin{equation}
\label{eq:s21}
	\left| S_{\si{21}}(f) \right|^2 = \left( 1 - \frac{\left( 1 - S^2_{\si{21,min}}\right)}{1 + \left( 2Q\frac{f - f_{\si{r}}}{f_{\si{r}}} \right)} \right)
.\end{equation}
 The $Q_i$ is determined from the equality $Q_{\si{i}} = Q/S_{\si{21,min}}$. The measured $1/Q_i$ can be expressed as:
 \begin{equation}\label{eq:tandelta}
 	\frac{1}{Q_{\si{i}}}(g) = p(g) \tan{\delta} + b(g)
 ,\end{equation}
 where $g$ makes it explicit that these quantities are dependent on the CPW geometry, $p(g)$ is the filling fraction of the dielectric film \cite{gao_physics_2008}, and $b(g)$ is the sum of all loss mechanisms other than the film's loss tangent. We calculated the $p(g)$ of each film by using the commercial solver Sonnet. The resulting $p(g)$ are displayed in Table \ref{tab:ff}.
 
 In the standard tunneling model (STM), the $\tan{\delta}$ is dependent on power, frequency, and temperature\cite{gao_physics_2008, Molina-Ruiz2020}:
 
 \begin{equation}\label{eq:neq}
 	\tan{\delta} = \tan{\delta_0} \tanh{\frac{\hbar\omega}{2k_B T}}\left( 1 + \frac{N}{N_0}\right)^{-\beta/2}
 ,\end{equation} 
 
 \noindent where $\tan{\delta_0}$ is the TLS-induced loss tangent at zero temperature and low internal resonator power, $N$ is the number of photons inside the resonator, $N_0$ is the critical photon number, and $\beta$ is equal to 1 in the STM, but has experimentally been found to range between 0.3 to 0.7 \cite{Molina-Ruiz2020}. The frequency $\omega$, the temperature $T$ , and the photon number $N$ are controlled by the experiment. We calculated $N$ from the internal resonator power \cite{Barends2009}.  
  
 We calculated an upper bound of $\tan{\delta}$ by setting $b(g)$ from Eq. \eqref{eq:tandelta} to zero. We also estimated $\tan{\delta}$ by making use of the $1/Q_{\si{i}}$ measurement of the reference chip on c-Si. In this case we set $b(g)$ equal to the $1/Q_{\si{i}}$ of the c-Si chip. 
 
\begin{figure} [ht]
\begin{center}
\begin{tabular}{c} 
\includegraphics[height=7.5cm]{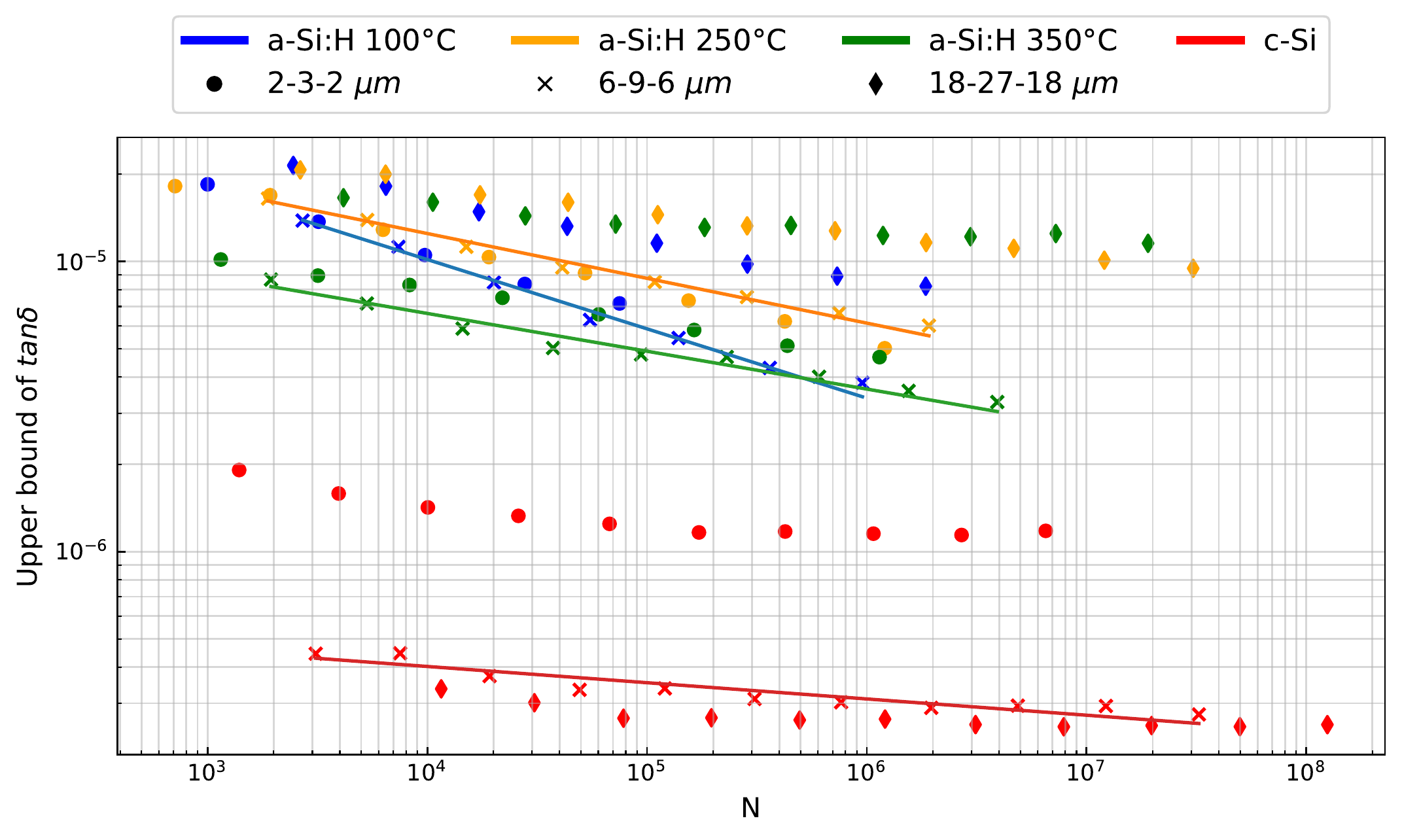}
\end{tabular}
\end{center}
\caption[Test]{\label{fig:tandeltalim}The upper bounds of the loss tangents of the three a-Si:H films and the c-Si reference chip, that were calculated from the measured $1/Q_{\si{i}}$. The photon number $N$ was calculated from the internal resonator power \cite{Barends2009}. The lines that connect the data points of the 6-9-6 $\si{\mu m}$ CPW resonators represent a fit to a power law and serve to roughly indicate the lowest measured values, and therefore the overall upper bounds of the materials' $\tan{\delta}$.} 
\end{figure} 
\begin{figure} [ht]
\begin{center}
\begin{tabular}{c} 
\includegraphics[height=7.5cm]{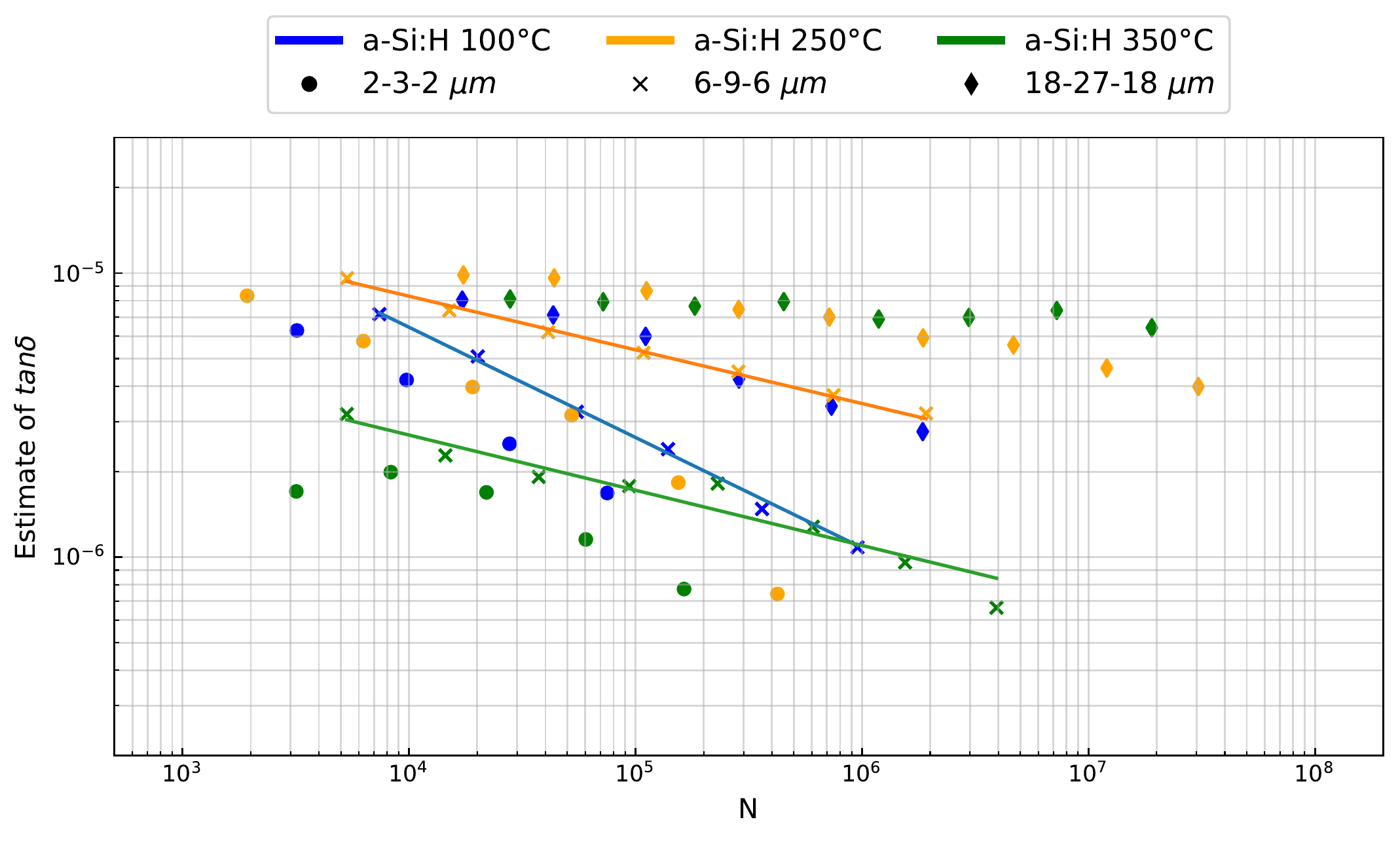}
\end{tabular}
\end{center}
\caption[Test]{\label{fig:tandeltaest}The estimates of $\tan{\delta}$ that were calculated from the measured resonator $1/Q_{\si{i}}$ by using the chip that was fabricated directly on top of the c-Si substrate as a reference. The photon number $N$ was calculated from the internal resonator power \cite{Barends2009}.} 
\end{figure} 

\subsection{Loss tangent results}
\noindent 
In Fig. \ref{fig:tandeltalim} we can see the upper bounds of the $\tan{\delta}$ of the three a-Si:H films, and of the reference chip that was fabricated directly on top of the c-Si substrate.  

 We observe that the a-Si:H films have an order of magnitude higher upper bound of $\tan{\delta}$ than the c-Si substrate. We do not observe a clear correlation between $T_{\si{sub}}$ and the upper bound of $\tan{\delta}$. The estimates of $\tan{\delta}$ that were calculated by using the c-Si chip as a reference are plotted in Fig. \ref{fig:tandeltaest}. We also do not observe a clear correlation between $T_{\si{sub}}$ and the estimate of $\tan{\delta}$. 

For the CPW resonators with the 6-9-6 $\si{\mu m}$ geometry we have added a fit to a power law. Looking at this fit, we observe that the 100 \degree C film has a steeper slope than the other two a-Si:H films. As can be seen in Eq. \eqref{eq:neq} this can be an indication of a difference in the $\beta$ parameter. 

Finally, we observe that for the a-Si:H films, the widest CPWs exhibit the largest $\tan{\delta}$, while the chip on c-Si exhibits the largest $\tan{\delta}$ for the narrowest CPWs. The plotted results have been corrected for the filling fractions that are listed in Table \ref{tab:ff}, as explained in Section \ref{sec:deter}.

\section{Discussion}
As expected, $T_{\si{sub}}$ strongly influences the structural and compositional properties of the materials. Increasing $T_{\si{sub}}$ results in films with less hydrogen, less voids, smaller microstructure parameter, less bond-angle disorder, and a higher infrared refractive index and therefore higher film density. Recent literature suggests that the dielectric loss is correlated to the dangling bond density $\rho_{\si{DB}}$ and not to the atomic silicon density, and that the dangling bond density decreases with $T_{\si{sub}}$\cite{Molina-Ruiz2020}. We recommend measuring the $\rho_{\si{DB}}$ of these materials using electron paramagnetic resonance spectroscopy to investigate its correlation with $\tan{\delta}$. 

Both for the upper bound of $\tan{\delta}$ and the estimate of $\tan{\delta}$ we do not observe a correlation with $T_{\si{sub}}$. However, because we lack data at lower photon numbers, we cannot draw conclusions about the low-power loss tangent $\tan{\delta_0}$: At the powers at which we measured, a variation in $\tan{\delta_0}$ cannot be distinguished from a variation in $N_0$. However, if we assume that the three a-Si:H films have the same $N_0$, then the data suggests a $\tan{\delta_0}$ that decreases monotonically with $T_{\si{sub}}$. 

We observe a dependence of $\tan{\delta}$ on the CPW geometry, even after accounting for the filling fractions of the a-Si:H films or the c-Si substrate. This is expected since the resonators have TLS-hosting interfacial layers \cite{gao_experimental_2008, Wenner2011, Woods2019, Barends2010}, whose filling fractions and therefore effect on the measured $\tan{\delta}$ are dependent on the CPW geometry. To gain  more insight into the origin of the dielectric loss it will be necessary to disentangle the contributions from surface TLSs and bulk TLSs.

The results at the power range in which we measured are directly applicable to MKIDs, that operate at relatively high powers in comparison with the mm-submm filters that operate at single photon energies. The a-Si:H films are promising for MKID detectors because all three films exhibit an excellent $\tan{\delta} < 10^{-5}$ at a resonator energy of $10^5$ photons, or at --55~dBm internal resonator power.

Although the frequency independent TLS density of states assumed by the STM makes us expect that the low microwave $\tan{\delta}$ will translate to a low $\tan{\delta}$ for mm-submm waves, we recommend measuring the $\tan{\delta}$ at mm-submm wavelengths. This would not only give us more information on the applicability of these films to mm-submm filters, but would also provide us with an interesting dataset to test one of the fundamental assumptions of the STM.

We want to point out that the analysis of $\tan{\delta}$ would be simplified by using microstrips instead of CPWs, thereby eliminating radiation losses and removing the influence from the substrate. This would also increase the sensitivity of the resonators to the $\tan{\delta}$ due to an increased filling fraction. Finally, we note that a practical benefit of the film that was deposited at 250 \degree C is that its stress $\sigma$ is almost zero, as can be seen in Appendix \ref{app:results}. 

\section{Conclusion}
Although the structural and compositional room temperature properties of the a-Si:H films show a clear dependence on $T_{\si{sub}}$, we do not observe a correlation of these properties with $\tan{\delta}$. All three a-Si:H films have a $\tan{\delta}$ below $\si{10^{-5}}$ at 120 mK and a resonator energy of $10^5$ photons. This makes these films promising for application in mm-submm on-chip filters. Further research at single-photon energies, at mm-submm wavelengths, and using microstrips or lumped element parallel plate capacitors is recommended. This would lead to more conclusive data on the possible differences between the materials' $\tan{\delta_0}$. Furthermore, we recommend to investigate if there is a correlation of the dielectric loss with the dangling bond density. 
\clearpage
\appendix
\section{PECVD deposition parameters}\label{app:depo}
\begin{table}[!htb]
\centering
\begin{longtable}{ll}
\caption[PECVD recipes.]{PECVD parameters.}\label{tab:recipes} \\
\endfirsthead
\toprule
 Material            & a-Si 100 / 250 / 350 \\ \midrule
 $T_\text{sub}$ (\degree C) & 100 / 250 / 350 \\ 
 RF power (W)            & 15                 \\ 
 \si{SiH_4} flow (sccm) &  25                 \\ 
 Ar flow (sccm) & 475                         \\  
 Pressure (Torr) & 1                          \\ 
 Deposition time (m$'$s$''$) & 10$'$0$''$ / 7$'$9$''$ / 7$'$0$''$       \\ 
 \bottomrule
\end{longtable}
\end{table}
\section{FTIR details}
Measurements were performed using a Thermo Fischer Nicolet, using a resolution of 4 \si{cm^{-1}}. Each measurement was performed after flushing the FTIR chamber with nitrogen for 15 minutes. We performed a baseline correction on the transmission data, that was always smaller than 2\% of the original transmission values, such that the resulting baseline transmission of the c-Si substrate equaled its theoretical value for a $n_{\si{ir}}$ of 3.42. 
\section{Bruggeman EMA host material details}
\label{app:EMA}
The host material that we used in the Bruggeman EMA to calculate the $f_{\si{v}}$ from the ellipsmetry data using CompleteEASE is the material \emph{a-Si parameterized} in CompleteEASE. This material is fully described by a Cody-Lorentz (CL) oscillator\cite{Patel2019} without ultraviolet and infrared poles, without an Urbach tail, and with the parameter values listed in Table \ref{tab:CL}.
\begin{table}[!htb]
\centering
\begin{longtable}{llllll}
\caption[Bruggeman EMA host material Cody-Lorentz parameters.]{Bruggeman EMA host material Cody-Lorentz parameters.}\label{tab:CL} \\
\endfirsthead
\toprule
 $A$ (eV) & $Br$ (eV)  & $E_0$ (eV) & $E_g$ (eV) & $E_p$ (eV) \\ \midrule
 121.874 & 2.547 & 3.594 & 1.583 & 1.943 \\ 

 \bottomrule
\end{longtable}
\end{table}

\section{Overview of the room temperature results}
\label{app:results}
\begin{longtable}{l|lll}
\caption{Overview of the results of the room temperature measurements.}
\endfirsthead
\toprule
Film & a-Si 100  & a-Si 250 & a-Si 350 \\ \midrule
          $T_\text{sub}$ (\degree C) & 100  & 250 & 350 \\ 
          $C_\text{H}$ (at.\%) & 22     & 17     & 13   \\ 
$R^*$         & 0.32    & 0.20     & 0.15    \\ 
$n_\text{ir}$           & 3.13     & 3.21     & 3.39  \\ 
$n_\text{ir}/n_\text{c-Si,ir}$    & 0.91     & 0.94     & 0.99 \\ 
$\Delta\theta$    & 11.12 $\pm$ 0.12     & 11.01 $\pm$ 0.11    & 10.48 $\pm$ 0.10  \\ 
$f_\text{v}$ (\%)    & 16.5   & 5.6     & 2.0  \\ 
$t_{\si{f}}$ (nm)    & 250.8 $\pm$ 0.1   & 236.0 $\pm$ 0.1    & 259.9 $\pm$ 0.1 \\
$t_{\si{n}}$ (nm)    & 5.8 $\pm$ 0.1   & 4.7 $\pm$ 0.1    & 4.2 $\pm$ 0.1 \\  
$\sigma$ (MPa)(tens.)   & 128.7     & 3.0     & -379.0 \\ 
\bottomrule
\end{longtable}
 \acknowledgments 
 The first author would like to express his gratitude towards Marco van der Krogt who has helped him with the deposition of the a-Si:H films. 
\bibliography{ms.bib} 
\bibliographystyle{spiebib} 

\end{document}